\documentstyle[prl,aps,preprint]{revtex}
\begin{document}
\title{Modelling an Imperfect Market}
\author{Raul Donangelo,\footnote{Permanent address: Instituto de F\'\i sica,
Universidade Federal do Rio de Janeiro, Cidade Universit{\'a}ria, CP 68528,
21945--970 Rio de Janeiro--RJ, Brazil}  
Alex Hansen,\footnote{Permanent address: Department of Physics, Norwegian 
University of Science and Technology, NTNU, N--7034 Trondheim, Norway}
Kim Sneppen \footnote{Permanent address: NORDITA, Blegdamsvej 17, DK-2100 
Copenhagen {\O}, Denmark}
and Sergio R.\ Souza$^*$}
\address{International Centre for Condensed Matter Physics,
Universidade de Bras{\'\i}lia, CP 04513, 70919--970 Bras{\'\i}lia--DF,
Brazil}
\address{and}
\address{NORDITA, Blegdamsvej 17, DK--2100 Copenhagen {\O}, Denmark}
\date{\today}
\maketitle
\begin{abstract} 
We propose a simple market model where agents trade
different types of products with each other by using money,
relying only on local information.
Value fluctuations of single products, combined with the condition
of maximum profit in transactions, readily lead to persistent
fluctuations in the wealth of individual agents.
\end{abstract} 
\section{Introduction}
\label{intro}
Financial markets usually consist in trades of commodities and currencies. 
However, one can easily find cases in other types of
human endeavour that parallel the activities observed
in financial markets.
For example, a politician may select the standpoints in his/her
platform in exchange for votes in the coming election\cite{D57}.
Also, scientists may trade ideas in order to generate citations.
However, the easiest quantifiable marketplace is the financial one,
where the existence of money allows a direct measure of value.

There have been several proposals to model such markets.
A recent review was presented by Farmer \cite{F99}.
Bak, Paczuski and Shubik \cite{BPS97} have proposed a model where they
consider the price fluctuations of a single product traded by many agents, 
all of which use the same strategy. 
In that model fat tails and anomalous Hurst exponents appear when global 
correlation between agents are introduced.
An alternative model, presented by Lux and Marchesi \cite{LM99},
is driven by exogenous fluctuations in ``fundamental'' values of
a single good, with induced non-gaussian fluctuations in price
assessment of this good arising from switching of strategies 
(trend followers and fundamentalists) by the agents. 

Below we give a few considerations that had led us in the development
of a model where the dynamics arises solely from the interaction of 
agents trying to exchange several types of goods.

Financial markets exhibit a dynamical behavior that, even in the absence
of production, allow people to become either wealthy or poor.
If in these markets there were some sort of equilibrium, e.g.\ due to
complete rationality of the players, this would not be possible.
In order to create wealth or bankruptcy, people have to outsmart each other.
This means that each trader attempts to buy as cheaply and sell as
expensively as possible.
This demands that an agent should find a seller which sells at a low price,
and later another one that is willing to buy the same product at a higher one.
Thus different agents price differently the same product.
In this work we demonstrate that it is possible to devise a simple model
for such a non-equilibrium market.
We call it the Fat Cat (FC) model, as it functions on greed
(each agent buys to optimize his own assets)
and creates a market with large fluctuations, i.e.\ fat tails. 
In Sect.\ \ref{model} we present this model and show examples of its
dynamical evolution. 
We show that it leads to an ever fluctuating market.
In the following Sect. \ref{fluctuations}, the time series generated by this
model is analyzed and it is shown that it exhibits 
persistency in the fluctuations of the wealth of individual agents. 
In the final section we summarize the work and present suggestions
for generalizing the model to let individual agents evolve their
individual trading strategies.
\section{Description of the FC model}
\label{model}
Consider a market with $N_{ag}$ agents, 
each having initially a stock of $N_{un}$ 
units of products selected among $N_{pr}$ different types of products.
In a previous work \cite{DS99} we have shown that, for the case of agents 
having memory of past transactions, such a market spontaneously selects
one of the products as the most adequate as a means of exchange. 
The product so chosen acts as money in the sense that it is accepted even when
the agent does not need it, because through the memory of past requests of
products the agent knows it is in high demand, and therefore it will be 
useful to have it to trade for other products.
The selection of a product as an accepted means of exchange is not indefinite:
after some time another one substitutes it as the favorite currency. 
The time scale for these currency substitutions is large when measured in 
number of exchanges between individual agents.
 
In the present model we consider that each of the agents initially has
$N_{mon}$ units of money.
According to the discussion in the preceding paragraph, in principle money
could be viewed as one of the products, but here we consider it a separate
entity in order to quantify prices.
The $N_{un}$ products given initially to each agent are selected
randomly.
Later, during the time evolution of the system, each agent $i$ 
has, at each time step, an amount of money $M (i)\,,i=1,\dots,N_{ag}$, and a 
stock of the different products $j$, $S (i,j)$, where $j=1,...,N_{pr}$.
Since the model uses money as means of exchange, agents assign different
prices to the different products in their possessions. 
The prices of the different items in the stock of agent $i$, $P(i,j)$, are
taken initially to be integers uniformly distributed in the interval $[1,5]$.  
We have verified that the evolution of the system does 
not depend on this particular choice.

How do we picture such a market?
We may imagine antique collectors trying to buy objects directly from each
other, using their own estimates for the prices of the different stock. 
When two agents meet, one of them, the buyer, checks the seller's price list,
and compares it with his own price list.  We have chosen the antique collector
market as an example because 
few other markets show spatial price fluctuations at 
such a high level.
The decision to buy or not, and the changes in the value of the agents'
products are given by some strategy, which for now assume is the same for
all agents in the system.
Among all the products that the buyer considers to be possible buys (having
a price set by the seller which is lower than the one he would sell the same
product for), he will single out the best one, and will then attempt to buy it.
But, if the buyer finds no products that he considers as good buys,
the seller will consider that he has overpriced his goods and will as a
consequence tend to lower his prices.
At the same time, the buyer will think that his price estimate was too low, 
and as a consequence raise his price estimate.

This is the basis for our computer simulation for such a market place.
In it, we assume that at each time step the following procedure takes place:
\begin{enumerate}
\item Buyer ($b$) and seller ($s$) are selected at random among the $N_{ag}$
agents.
If the seller has no products to offer, then another seller is chosen.\\

\item The buyer selects the product $j$ in the seller's stock which maximizes
$P(b,j)-P(s,j)$ (i.e. his profit).
The corresponding $j$, we call $j_{bb}$ (best buy).\\

\item If the buyer does not have enough money, 
(i.e.\ if $M(b) < P(b,j_{bb})$), we go back to the first step, choosing
a new pair of agents.\\

\item If the buyer has enough money we proceed.\\
If $P(s,j_{bb}) < P(b,j_{bb})$, the transaction is performed at the
seller's price. This means that we adjust:
$S(b,j_{bb})\rightarrow S(b,j_{bb})+1$, $S(s,j_{bb})\rightarrow S(s,j_{bb})-1$,

$M(b)\rightarrow M(b)-P(s,j_{bb})$, $M(s)\rightarrow M(s)+P(s,j_{bb})$.\\

\item If $P(s,j_{bb})\ge P(b,j_{bb})$, the transaction is not performed.\\
In this case, the seller lowers his price by one unit,

$P(s,j_{bb})\rightarrow \max(P(s,j_{bb})-1,0)$,

and the buyer raises his price by one unit,

$P(b,j_{bb})\rightarrow P(b,j_{bb})+1$.\\
\end{enumerate}

We see that, according to these rules, buyer and seller decide on a
transaction based only on their local information, i.e. their
estimates of the prices for the different products they possess.
These prices are always non-negative integers.
Also note that since, as defined in step 3 above, the price offered by the
buyer cannot be higher than the amount of money he has, we are not
allowing for the agents to get in debt. 
Further, the model tends to equilibrize large price differences, according
to step 5, but induces price differences when buyer and seller agree on
the price of the most tradeable product.
This non-equilibrizing step is essential to induce dynamics in a model like
the present one, where all agents follow precisely the same strategy.
Without it the system would freeze into a state where all agents agree on all
prices. 

The rules given above are just one possible set of rules for transactions.
We have found other sets that lead to a behavior qualitatively similar
to the one shown below for the present rules.
\footnote{A quote from Marx is appropriate here: ``These are my principles.  
If you do not like them, I have others" \cite{M34}.}
We now show that, under this set of local rules, the distribution of wealth
organizes itself into a dynamically stable pattern, and the same phenomenon
takes place with the prices.

One should emphasize that there is not an accepted market value for the
products.
Indeed, due to the price adjustments performed in unsuccessful encounters, 
the prices never reach equilibrium, and different agents may assign
different prices for the same product.
In Fig.\ \ref{fig1} we illustrate this point, showing the price assigned
by two different agents to the same product, as a function of time.
Time is defined here in terms of the number of encounters between agents,
and one time unit is the average time between events where a given agent
acts as a buyer.
We note that during a considerable fraction of the time there is a relatively
large difference between the prices assigned by the agents.
This shows that there is a margin for making profit in such a market,
i.e.\ arbitrage is possible. We have verified that
the average market price of a good
fluctuates with a Hurst exponent of $\sim 0.5$.

In Fig.\ \ref{fig2} we show, for the same time interval, an example of the
evolution of key quantities in the model associated to Agent 1 in
Fig.\ \ref{fig1}.
The total wealth of an agent $i$ is the amount of money plus the 
value of all goods in the agent's possession:
\begin{equation}
\label{eq1}
w(i) = M(i) + G(i)\;.
\end{equation}
Here the value of product $j$ is defined as the average of what all
agents consider its value to be:
\begin{equation}
\label{eq2}
P_{ave} (j)  =  \frac{1}{N_{ag}} \sum_{i=1}^{N_{ag}} P(i,j)\;,
\end{equation}
and the value of all agent $i$'s goods $G(i)$ is then defined as
\begin{equation}
\label{eq3}
G(i)=\sum_j S(i,j) P_{ave} (j)\;.
\end{equation}

We note that there are considerable fluctuations in the wealth of this
agent. The study of these fluctuations is essential to the understanding
of the properties of the model, and we develop this in the following section.

\section{Fluctuations in the FC model}
\label{fluctuations}

In order to quantify the fluctuations in wealth, we show in Fig.\ \ref{fig3}
the RMS fluctuations of the wealth of a selected agent as function of time. 
The figure illustrates that the wealth fluctuations can be characterized
by a Hurst exponent \cite{F88} $H\approx 0.7 $. 
We have examined variants of both model parameters and rules
to check the stability of this result.
We found it to be stable, as long as one keeps greed in the model.
For example, if one reduces the number of product types to only 
$N_{pr}=2$ (keeping $N_{ag}=100$, $N_{mon}=500$ and $N_{un}=100$ initially)
the Hurst exponent remains unchanged although the scaling regime shrinks.
Similarly setting $N_{un}=2$ (with $N_{pr}=100$
$N_{ag}=100$ and $N_{mon}=500$) resembling antique dealing where
each agents owns a few of many possible products, also
lets the Hurst exponent unchanged.
On the other hand, if greed is removed from the model, e.g.\ 
the buyer selects a product at random from the sellers store,
without consideration to the profit margin, the Hurst exponent
drops to $0.5$, signaling that no correlations develop in such a case.

Thus, the optimization of product selection expressed by step 2 in the
procedure is closely related to the persistent fluctuations seen in our model.
We have checked that other optimization procedures, as e.g.\ selecting the
cheapest product or the product the buyer has the least of in stock also give
similar persistent fluctuations.
Oppositely, random selection reflect an economy where different products do
not interact significantly with each other, and where our market of $N_{pr}$
different types of products nearly decouples into $N_{pr}$ different markets.
With random selection, the only interaction between products is indirect;
it appears due to the constraint of the agents' money take only non-negative
values. Overall, the random strategy gives a less 
fluctuating market where agents agree more on prices.
Greed indeed makes our model world both richer and more interesting 
(which is {\it not\/} to say {\it better.\/}).

We now try to quantify these wealth fluctuations. Fig.\ \ref{fig4}
displays the changes in the value of $w$ for several time intervals $\Delta t$.
The three curves are histograms of wealth changes for respectively 
$\Delta t=10$,   $\Delta t=100$ and   $\Delta t=1000$.
One observes fairly broad distributions with a tendency to asymmetry
in having bigger probability for large losses than for large gains.
Similar skewness is seen in real stock market data.
Furthermore, the tails are outside the Gaussian regime \cite{Plerou99}. 
In the upper panel of Fig.\ \ref{fig5}, this is investigated further by
plotting the histograms as function of the logarithmic changes
in $w$, $r_{\Delta t}=\log_2 w(t+\Delta_t)-\log_2 w(t)$ ({\sl log-returns}),
for the same three time intervals.
In that figure we have collapsed the curves onto each other by
rescaling them using the Hurst exponent $H=0.69$, consistent with  
the one found in Fig.\ \ref{fig3}. For the two short time intervals
the collapse is nearly perfect, even in the non-Gaussian fat tails.
For larger time intervals the distribution changes from a steep power 
law or stretched exponential, to an exponential, and finally becomes
Gaussian for very large intervals (not shown as it is very narrow on the
scale of this figure). 
We think it is interesting to notice that our model is consistent with
the empirical observation that in real markets the probability 
for large negative fluctuations is larger than that for large positive ones.

In the lower panel of Fig. \ref{fig5} we examine in details the fluctuations
for $\Delta t=1$, and compare the fat tails with truncated power law decays
$P(r) \sim 1/r^4 \cdot \exp(-|r|/R)$ which for such small time intervals
is nearly symmetrical.
The $1/r^4$ is consistent with the fat tail observed on 5-minute interval
trading of stocks \cite{Plerou99}.
The cut off size $R=0.8$
corresponds to cut offs when price changes are about
a factor 2 from the original price, a regime which is not addresses in the
short time trading analysis of \cite{Plerou99}.
We stress that our analysis of fat tails includes a wide distribution
of wealth, thus large relative changes of wealth 
are presumably mostly associated to poor agents.
Thus the seemingly good fit to fat tails observed
for stock market fluctuations in the 1000 largest US companies
\cite{Plerou99} may be coincidental.

\section{Summary and Discussion}
\label{theend}

The appearance of fat tails
\cite{M63,E95,Plerou99} and Hurst exponents\cite{E95,M91} 
larger than 0.5 in the
distribution of monetary value appears to be a characteristic of
real markets.
The present model is, as it was the case with 
the previous version \cite{DS99}, qualitatively consistent with these features. 
We stress that here, as for the previous
model, we are not including any development of strategy by the agents that
might force the emergence of cooperativity \cite{Z97,PB99}.  
A more important difference to game theoretic models is that the minority game,
as well as the evolving Boolean network of Paczuski, Bassler and Corral
\cite{PB99} evolves on basis on a global reward function.
With the present model we would like to open for models which evolve
with imperfect information, 
preferably in a form which allows direct
comparison with financial data.
The present model does this, and in a setting where there are many products
and thus possibilities for making arbitrage along different ``coordinates''.

Compared to models with fat tails or persistency arising from 
boom-burst cycles, as the trend enhancing model
of Delong \cite{Delong90} or the trend following model of Lux and Marchesi
\cite{LM99}, the present model discuss 
anomalous scaling in a market where no agent has a precise knowledge of the 
global or average value of a product.
There is only local optimization of utility (estimated market value) and all
trades are done locally without the effects of a global information pool.
This imperfect information gives a possibility for arbitrage and opens for 
a dynamic and evolving market. 

The model we propose is for a market composed of agents, goods and money. 
We have demonstrated that such a market easily shows persistent
fluctuations of wealth, and seen that this persistency is closely related 
to having a market with several products which influence each others´ trading.
As seen in Fig.\ \ref{fig2}, wealth increase of an agent is associated
with active trading of few products.  This may be understood as
follows:  when the number of options a seller presents is small,
it is hard for the buyer to find a bargain, so if a transaction is
performed it will probably be at a good price for the seller.

The simpler model of Ref.~\cite{DS99}, had persistency in the fluctuations in
the demand for different products, whereas, as mentioned above, the persistence
here is in the fluctuations in the wealth of the different agents.
It is interesting to mention that the Hurst exponents in the two models take
very similar values, and this for a wide variety of the respective parameters.
There is a kind of duality in that in the simpler model a product increased
in value when it was held by relatively few agents, whereas in the present
model the agents increase their wealth by specializing to few products.

The setup proposed here with agents and products with individual local
prices allows for also different individual strategies of the agents. 
For example both the selection of greed versus random strategy under 
step 2, and the particular adjustment of values defined under step 4-5
could be defined differently from agent to agent.
One may accordingly have different strategies for different agents, and each
agent could change its strategy in order to improve his performance. The
evolution of these strategies would then become an inherent part of the
dynamics. This opens for evolution of strategies as part of the 
financial market, and will be discussed in a separate publication\cite{DHSS}.
\vspace{0.5cm}

We thank F.A.\ Oliveira and H.N.\ Nazareno for warm
hospitality and the I.C.C.M.P.\ 
for support during our stay in Bras\'i lia.  
R.D., A.H.\ and S.R.S\ thank NORDITA for support during our stay in 
Copenhagen.  R.D.\ and S.R.S.\ thank MCT/FINEP/CNPq (PRONEX) under 
contract number 41.96.0886.00 for partial financial support.

\begin{figure}
\caption{Selling prices of a specific product by two different agents, as a function
of time. The simulation was performed for the following parameter values:
$N_{ag}=100$, $N_{pr}=100$, $N_{mon}=500$ and $N_{un}=100$, which apply to the
calculations presented in all figures in this work.}
\label{fig1} 
\end{figure}

\begin{figure}
\caption{Amount of money $M$, number of different products and combined wealth
$w=M+G$, held by an agent as a function of time.
See text for further
details.}
\label{fig2}
\end{figure}

\begin{figure}
\caption{ RMS fluctuations of the wealth of a given agent as function of time
interval $\Delta t$. In order to guide the eye we also plot the power
function $\Delta t^{0.7}$.}
\label{fig3}
\end{figure}

\begin{figure}
\caption{Probability of having changes in wealth
$\Delta w=w(t+\Delta t)-w(t)$ as a function of their size,
for the three different time steps
$\Delta t =10$ (full line),
$\Delta t = 100$ (long dashed line), and
$\Delta t = 1000$ (short dashed line).}
\label{fig4}
\end{figure}

\begin{figure}
\caption{a) Similar to Fig.~4 except that wealth fluctuation is measured here
in units of $\log_2 \left[ w(t+\Delta t)/w(t) \right] / \Delta t^H$,
where $\Delta t$ takes the same values as above, and we took for the Hurst
exponent the value $H = 0.69$.
b) Fits to the probability of having a wealth loss (gain) as a function
of the log-returns $r=\log_2 \left[ w(t+\Delta t)/w(t) \right]$, for
the case $\Delta t=1$. The fit 
$P(r) = 5/(r^2+0.03)^2 \cdot \exp(-|r|/0.4)$
have asymptotic expressions for large $|r|$ of the form
$P(r) \sim 1/r^4 \cdot \exp(-|r|/R)$, with $R=0.4$.}
\label{fig5}
\end{figure}
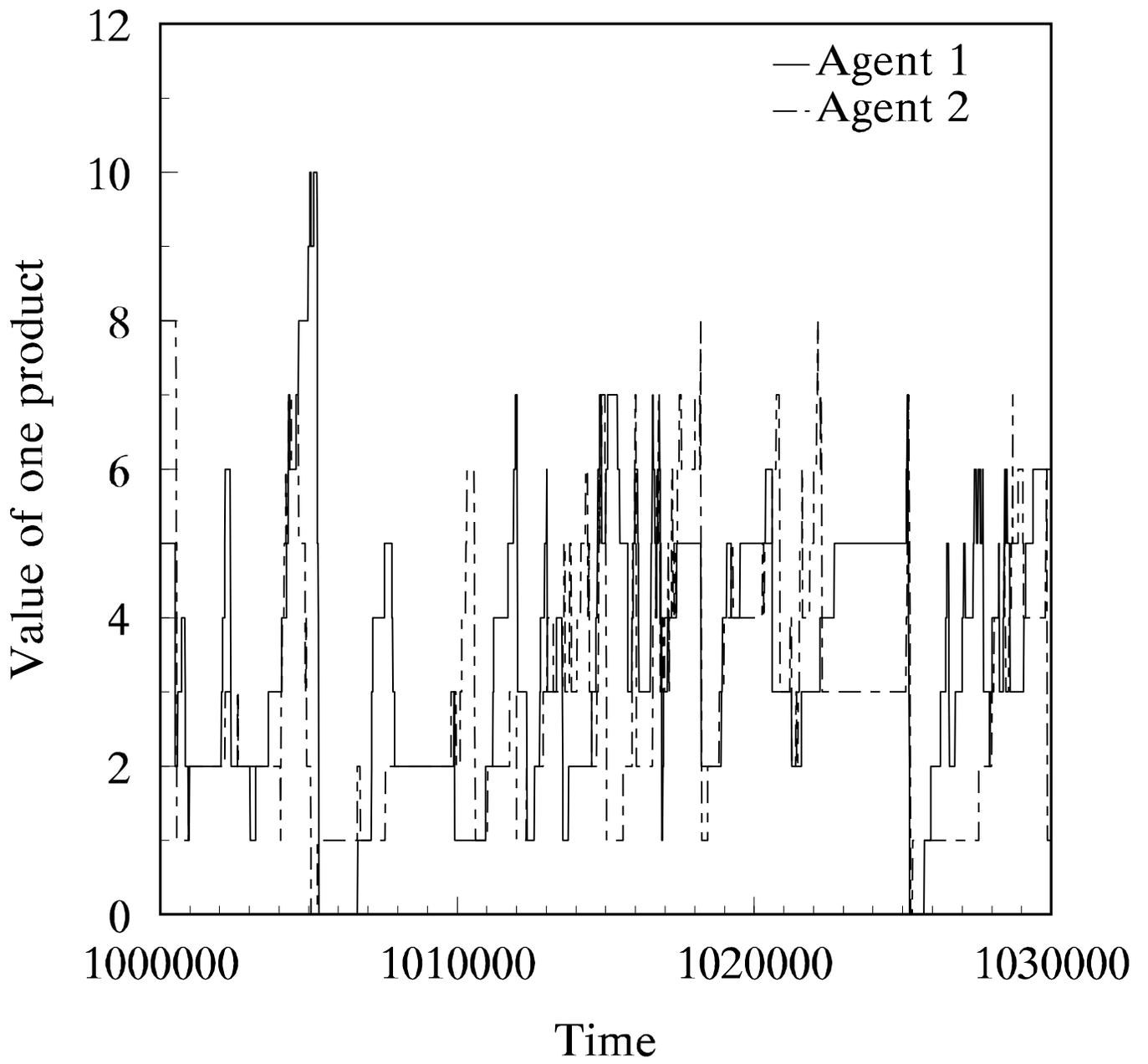
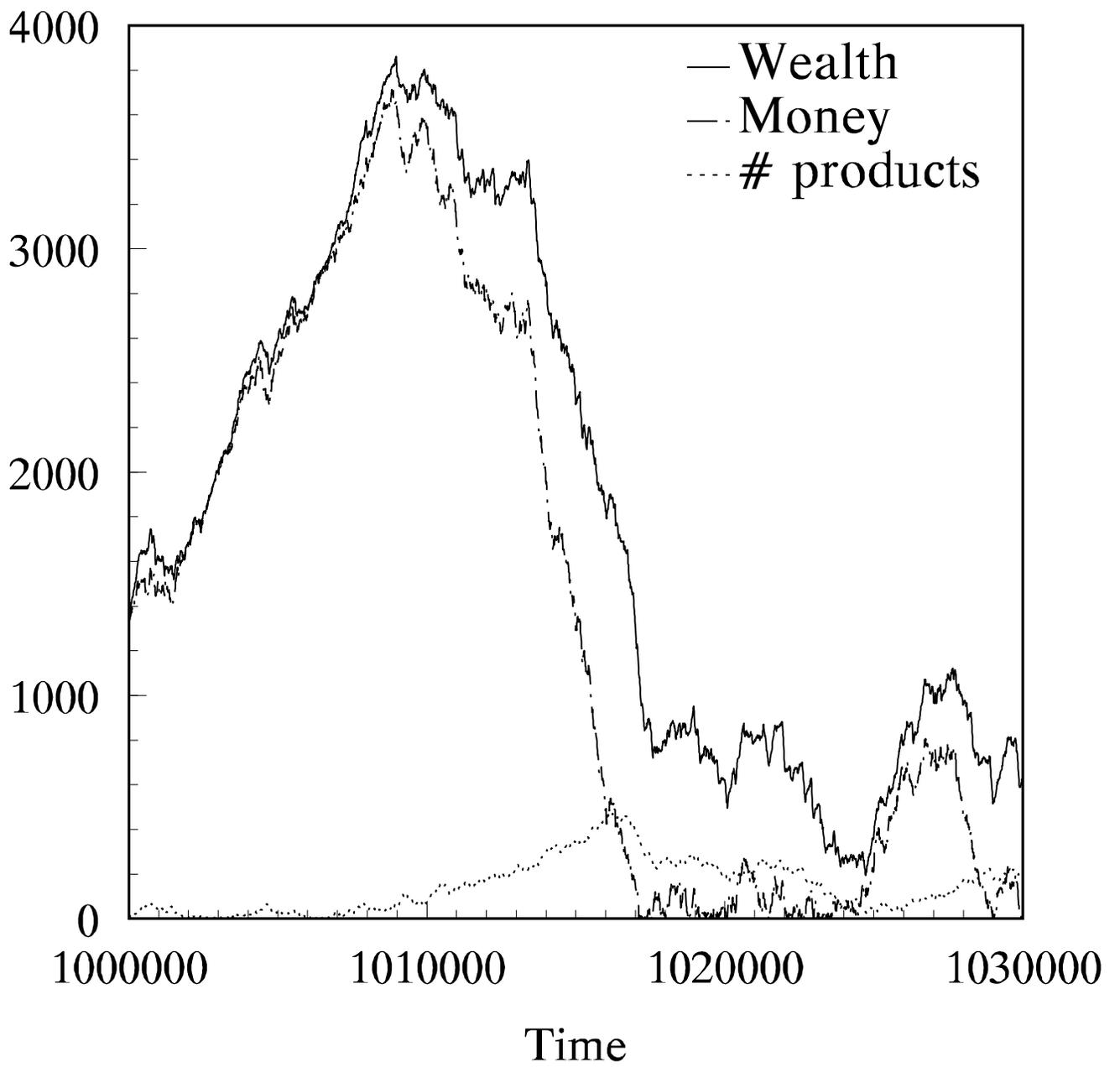
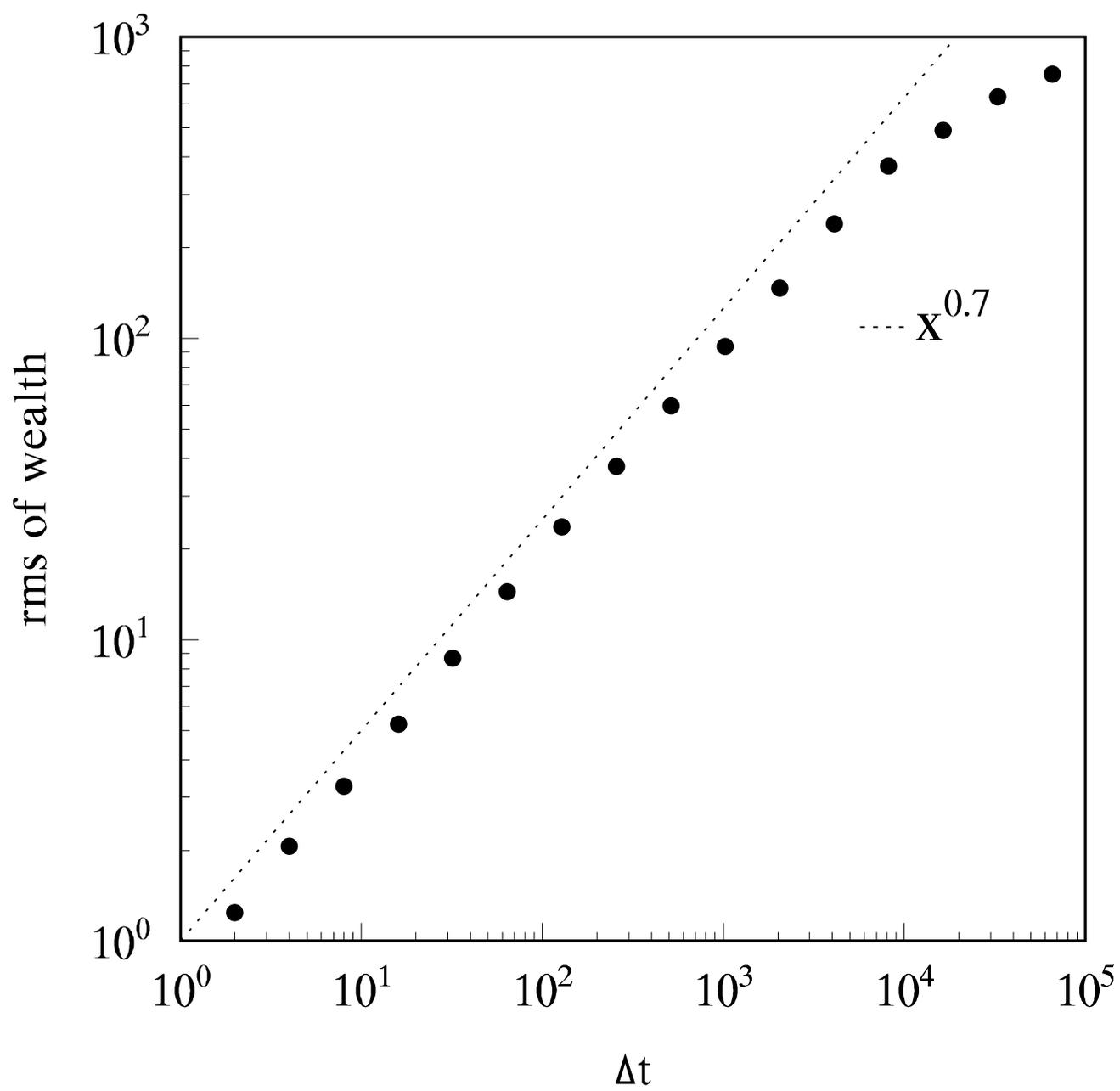
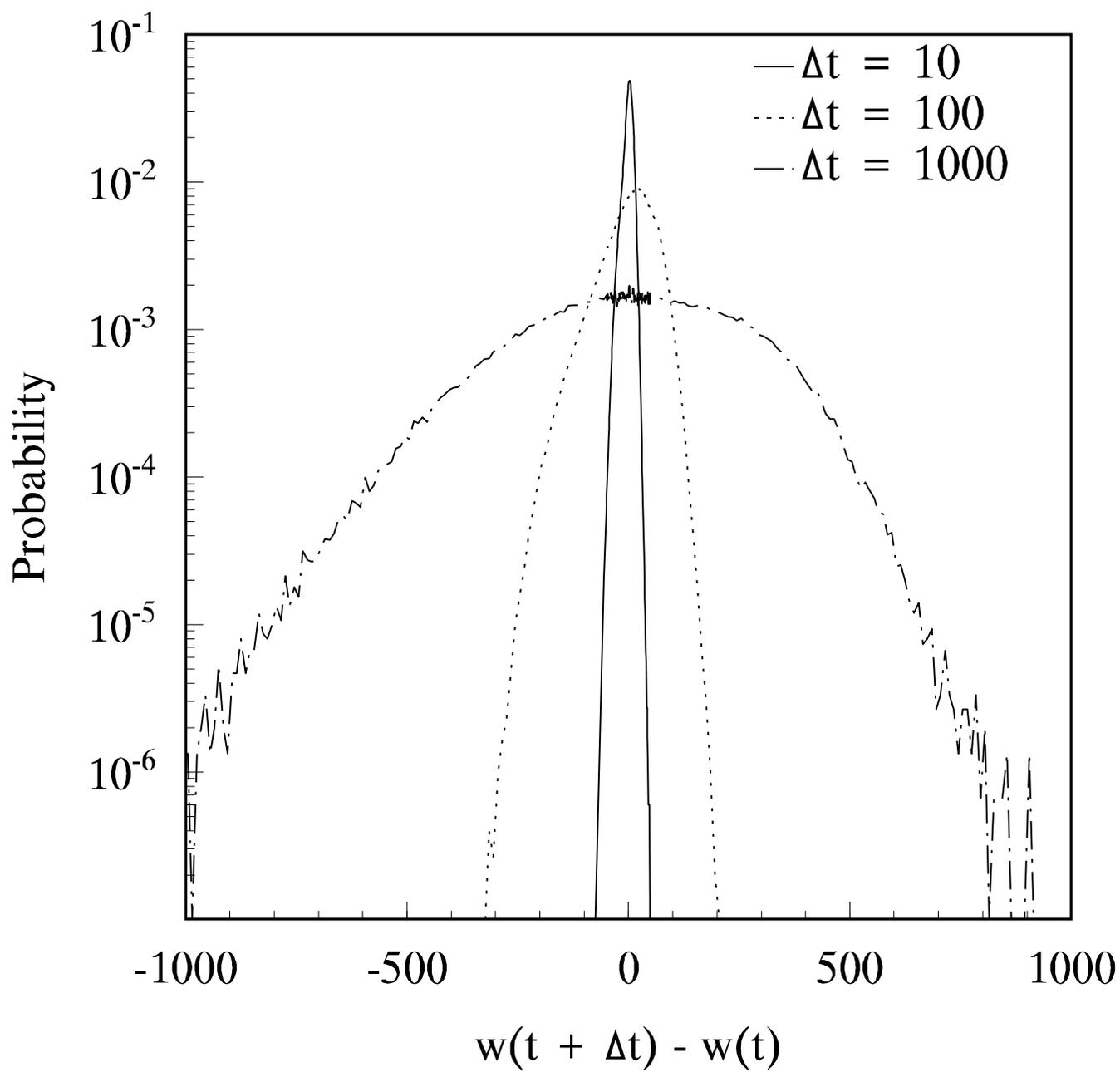
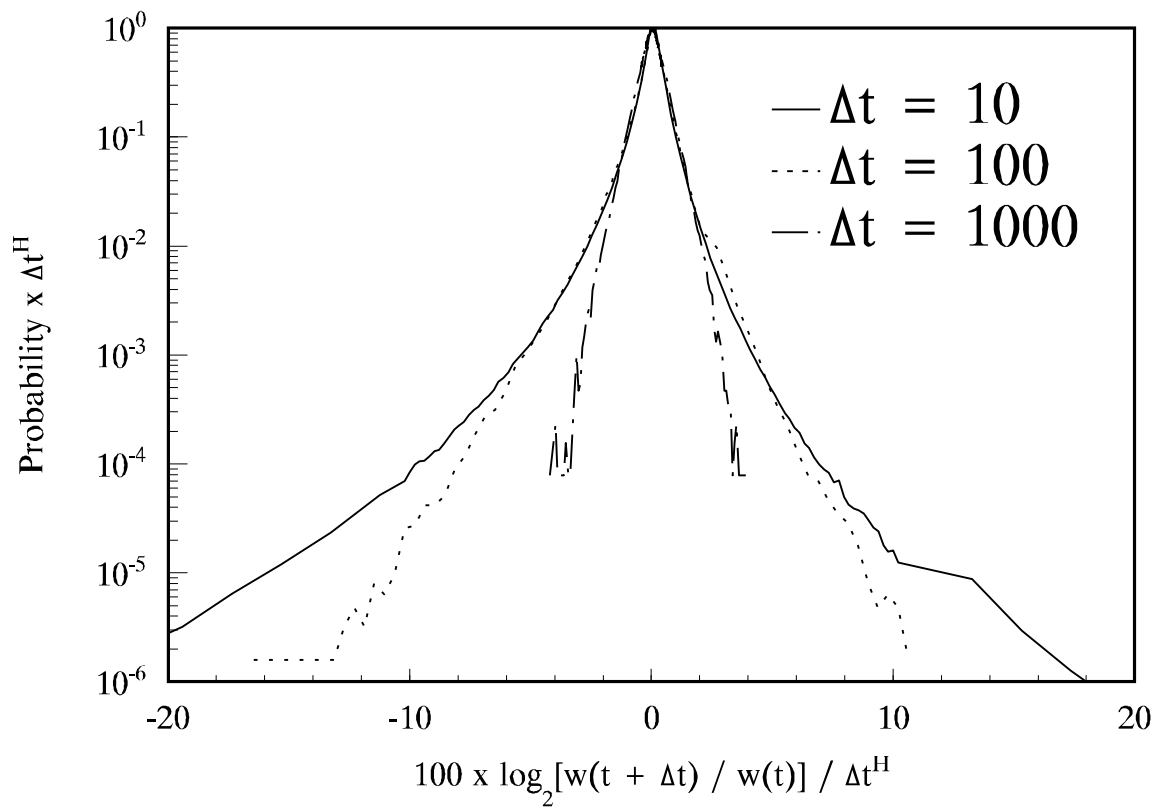
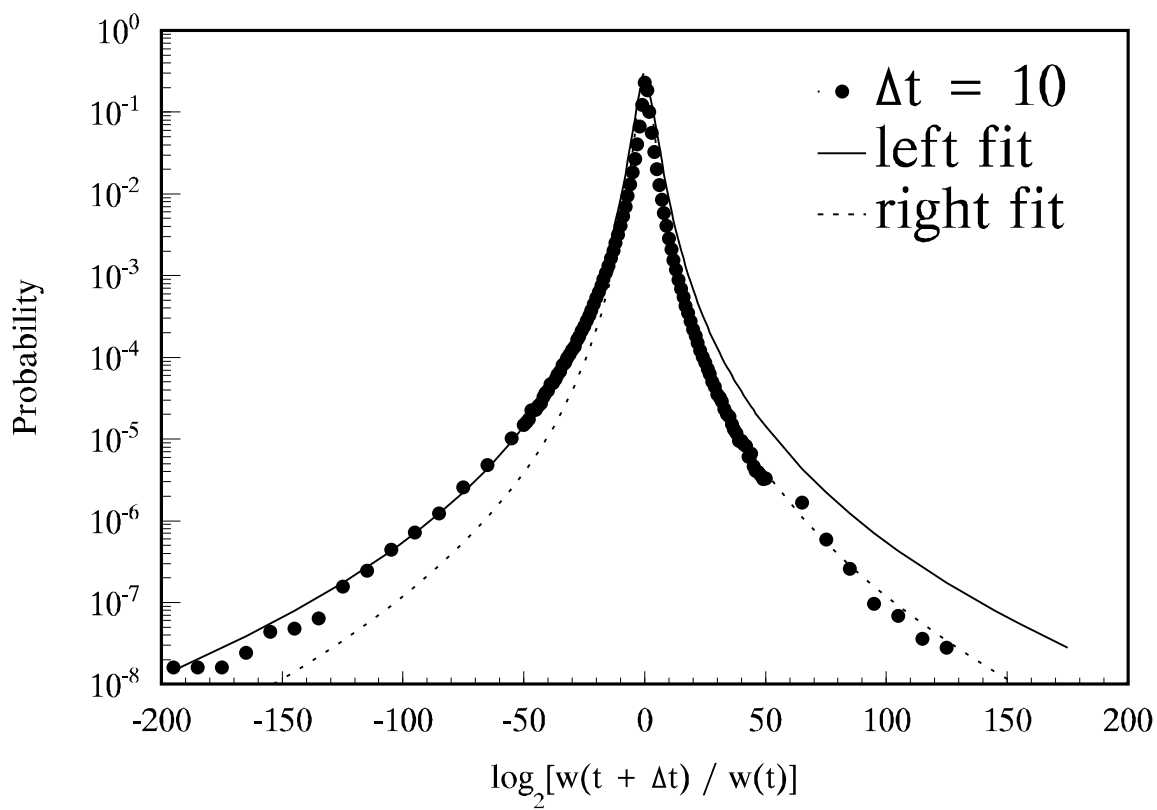
\end{document}